\documentclass[twocolumn,amsmath,amssymb,floatfix,pfra,aps]{revtex4}
\usepackage{graphicx}
\begin{document}
\title{Synthetic magnetic fields for ultracold neutral atoms}
\author{Y.-J.~Lin$^1$}
\author{R.~L.~Compton$^1$}
\author{K.~Jim$\acute{\rm e}$nez-Garc$\acute{\rm i}$a$^{1,2}$}
\author{J.~V.~Porto$^1$}
\author{I.~B.~Spielman$^1$}

\affiliation{$^1$Joint Quantum Institute, National Institute of
Standards and Technology, and University of Maryland, Gaithersburg,
Maryland, 20899, USA} \affiliation{$^2$Departamento de F\'{\i}sica,
Centro de Investigaci\'{o}n y Estudios Avanzados del Instituto
Polit\'{e}cnico Nacional, M\'{e}xico D.F., 07360, M\'{e}xico}

\date{\today}

\begin{abstract}
Neutral atomic Bose condensates and degenerate Fermi gases have been
used to realize important many-body phenomena in their most simple
and essential forms\cite{Greiner03,Regal04,Zwierlein04}, without
many of the complexities usually associated with material systems.
However, the charge neutrality of these systems presents an apparent
limitation - a wide range of intriguing phenomena arise from the
Lorentz force for charged particles in a magnetic field, such as the
fractional quantum Hall states in two-dimensional electron
systems\cite{tsui82,laughlin83}. The limitation can be circumvented
by exploiting the equivalence of the Lorentz force and the Coriolis
force to create synthetic magnetic fields in rotating neutral
systems. This was demonstrated by the appearance of quantized
vortices in pioneering
experiments\cite{Zwierlein05,schweikhard04,madison00,Aboshaeer01} on
rotating quantum gases, a hallmark of superfluids or superconductors
in a magnetic field. However, because of technical issues limiting
the maximum rotation velocity, the metastable nature of the rotating
state and the difficulty of applying stable rotating optical
lattices, rotational approaches are not able to reach the large
fields required for quantum Hall
physics\cite{juzeliunas04,Jaksch03,sorensen05}. Here, we
experimentally realize an optically synthesized magnetic field for
ultracold neutral atoms, made evident from the appearance of
vortices in our Bose-Einstein condensate. Our approach uses a
spatially-dependent optical coupling between internal states of the
atoms, yielding a Berry's phase\cite{Berry84} sufficient to create
large synthetic magnetic fields, and is not subject to the
limitations of rotating systems; with a suitable lattice
configuration, it should be possible to reach the quantum Hall
regime, potentially enabling studies of topological quantum
computation.
\end{abstract}

\maketitle

In classical electromagnetism, the Lorentz force for a particle of
charge $q$ moving with velocity ${\bf v}$ in a magnetic field ${\bf
B}$ is ${\bf v}\times q{\bf B}$. In the Hamiltonian formulation of
quantum mechanics, where potentials play a more central role than
fields, the single-particle Hamiltonian is ${\cal H} =\hbar^2({\bf
k}-q{\bf A}/\hbar)^2/2m$, where ${\bf A}$ is the vector potential
giving rise to the field ${\bf B}=\nabla \times {\bf A}$, $\hbar
{\bf k}$ is the canonical momentum and $m$ is the mass. In both
formalisms, only the products $q{\bf B}$ and $q{\bf A}$ are
important. To generate a synthetic magnetic field ${\bf B}^*$ for
neutral atoms, we engineer a Hamiltonian with a spatially dependent
vector potential ${\bf A}^*$ producing ${\bf B}^*=\nabla \times {\bf
A}^*$.

The quantum mechanical phase is the relevant and significant
quantity for charged particles in magnetic fields. A particle of
charge $q$ traveling along a closed loop acquires a phase $\phi=2\pi
\Phi_{B}/\Phi_{0}$ due to the presence of magnetic field ${\bf B}$,
where $\Phi_{B}$ is the enclosed magnetic flux and $\Phi_{0}=h/q$ is
the flux quantum. A similar path-dependent phase, the Berry's
phase\cite{Berry84}, is the geometric phase acquired by a slowly
moving particle adiabatically traversing a closed path in a
Hamiltonian with position dependent parameters. The Berry's phase
depends only on the geometry of the parameters along the path, and
is distinct from the dynamic contribution to the phase which depends
upon the speed of the motion.

The close analogy with the Berry's phase implies that properly
designed position-dependent Hamiltonians for neutral particles can
simulate the effect of magnetic fields on charged particles. We
create such a spatially-varying Hamiltonian for ultracold atoms by
dressing them in an optical field that couples different spin
states. The appropriate spatial dependence can originate from the
laser beams' profile\cite{juzeliunas04,juzeliunas06,gunter09} or, as
here, a spatially-dependent laser-atom detuning\cite{Spielman09}. An
advantage of this optical approach compared to rotating gases is
that the synthetic field exists at rest in the lab frame, allowing
all trapping potentials to be time-independent.

The large synthetic magnetic fields accessible by this approach make
possible the study of unexplored bosonic quantum-Hall states,
labeled by the filling factor $\nu= \Phi_{B}/\Phi_{0}$, the ratio of
atom number to the number of flux quanta.  The most outstanding open
questions in quantum-Hall physics center on states whose elementary
quasiparticle excitations are anyons: neither bosons nor fermions.
In some cases these anyons may be {\it non-abelian}, meaning that
moving them about each other can implement quantum gates, thus
non-abelian anyons are of great interest for this ``topological''
quantum computation\cite{nayak08}.  In electronic systems, the
observed $\nu=5/2$ quantum-Hall state may be such a system, but its
true nature is still uncertain\cite{Radu08}.  In contrast, the
$\nu=1$ bosonic quantum-Hall state with contact interactions has the
same non-abelian anyonic excitations as the $ \nu=5/2$ state in
electronic systems is hoped to\cite{cooper08}.

\begin{figure*}
\includegraphics[width=160 mm]{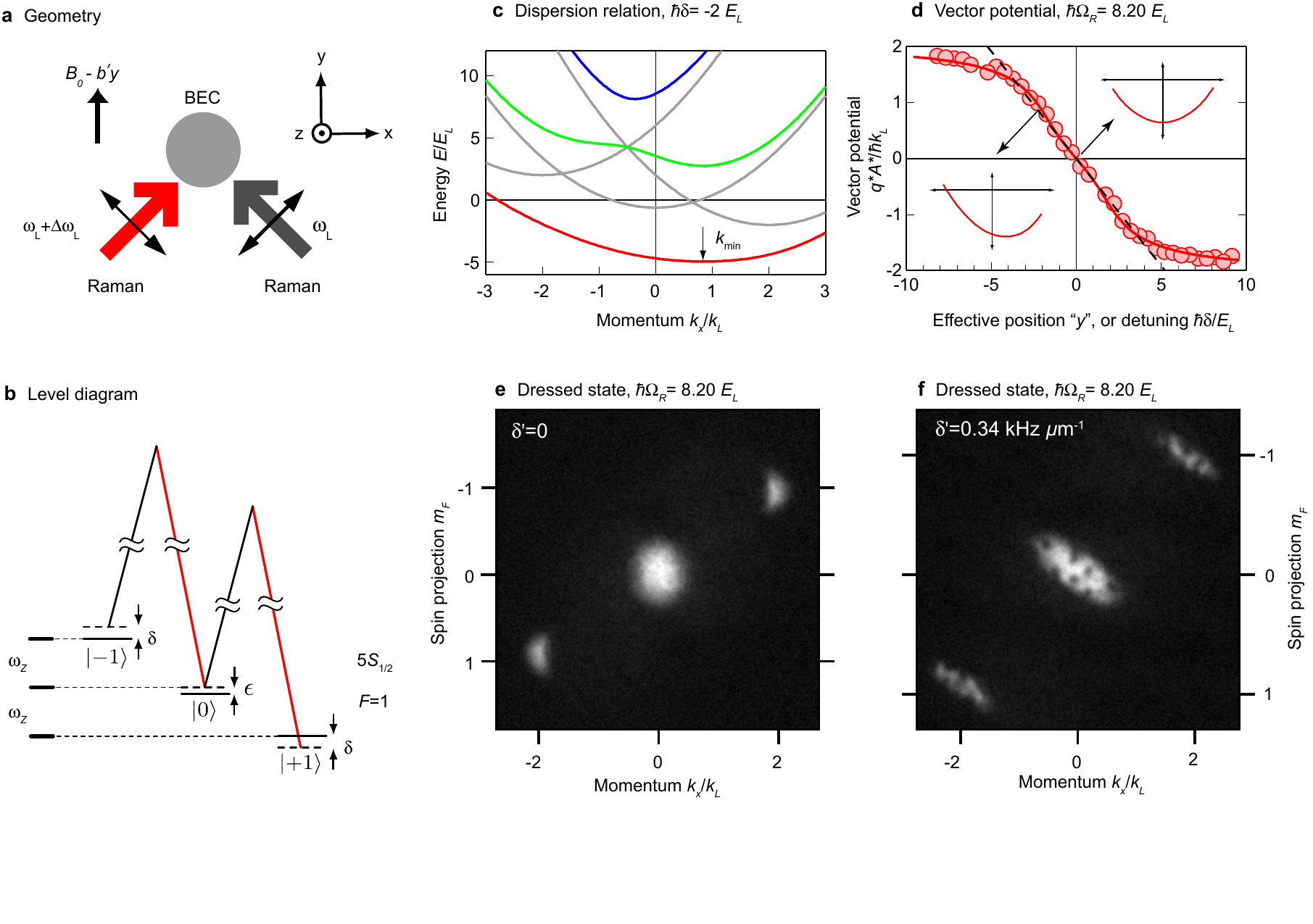}
\caption{{\bf Experiment summary for synthesizing magnetic fields.}
{\bf a}, The BEC is in a crossed dipole trap in a magnetic field
${\bf B}=(B_{0}-b^{\prime}y)\hat{y}$. Two Raman beams propagating
along $\hat{y}\mp\hat{x}$ (linearly polarized along
$\hat{y}\pm\hat{x}$) have frequencies $\omega_L$, $\omega_{L}+\Delta
\omega_{L}$. {\bf b}, Raman coupling scheme within the $F=1$
manifold: $\omega_{Z}$ and $\epsilon$ are the linear and quadratic
Zeeman shifts, and $\delta$ is the Raman detuning. {\bf c},
Energy-momentum dispersion relations. The grey curves represent the
states absent Raman coupling; the three colored curves are
$E_j(k_x)$ of the dressed states. The arrow indicates the minimum at
$k_{\rm{min}}$. {\bf d}, Vector potential $q^*A_x^*=\hbar
k_{\rm{min}}$ versus Raman detuning $\delta$. The insets show the
dispersion $E_1(k_x)$ for $\hbar \delta=0$ (top inset) and $-2E_L$
(bottom inset). {\bf e},{\bf f}, Dressed BEC imaged after a 25.1~ms
TOF without (e) and with (f) a gradient. The spin components,
$m_F=0,\pm 1$, separate along $\hat{y}$ due to the Stern-Gerlach
effect.} \label{setup}
\end{figure*}

To engineer a vector potential ${\bf A}^*=A^*_x\hat{x}$, we
illuminate a $^{87}$Rb BEC with a pair of Raman laser beams with
momentum difference along $\hat{x}$ (Fig.~1a). These couple together
the three spin states, $m_F=0$ and $\pm 1$, of the $5 S_{1/2}, F=1$
electronic ground state (Fig.~1b), producing three dressed states
whose energy-momentum dispersion relations $E_{j}(k_x)$ are
experimentally tunable. Example dispersions are illustrated in
Fig.~1c. The lowest of these, with minimum at $k_{\rm {min}}$,
corresponds to a term in the Hamiltonian associated with the motion
along $\hat{x}$, ${\cal H}^*_x \approx
\hbar^2(k_x-k_{\rm{min}})^2/2m^{*}=\hbar^{2}(k_x-q^*A_{x}^*/\hbar)^2/2m^*$,
where $A^*_x$ is an engineered vector potential that depends on an
externally controlled Zeeman shift for the atom with a synthetic
charge $q^*$, and $m^*$ is the effective mass along $\hat{x}$. To
produce the desired spatially-dependent $A^*_x(y)$ (Fig.~1d),
generating $-B^*\hat{z}=\nabla \times {\bf A}^*$, we apply a Zeeman
shift that varies linearly along $\hat{y}$. The resulting $B^*$ is
approximately uniform near $y=0$, at which point $A^*_x=B^*y$.
(Here, the microscopic origin of the synthetic Lorentz
force\cite{Cheneau08} is optical along $\hat{x}$, depending upon the
velocity along $\hat{y}$; the force along $\hat y$ is magnetic,
depending upon the $\hat x$ velocity.) By these means, we engineer a
Hamiltonian for ultracold atoms, that explicitly contains a
synthetic magnetic field, with vortices in the ground state of a
BEC. This is distinctly different from all existing experiments,
where vortices are generated by phase
imprinting\cite{Leanhardt02,andersen06},
rotation\cite{schweikhard04,madison00,Aboshaeer01}, or a combination
thereof\cite{matthews99}. Each of these earlier works presents a
different means to impart angular momentum to the system yielding
rotation. Fig.~1e shows an experimental image of the atoms with
$B^*=0$. Fig.~1f, with $B^*>0$, shows vortices. This demonstrates an
observation of an optically induced synthetic magnetic field.

\begin{figure*}
\begin{center}
\includegraphics[width=160 mm]{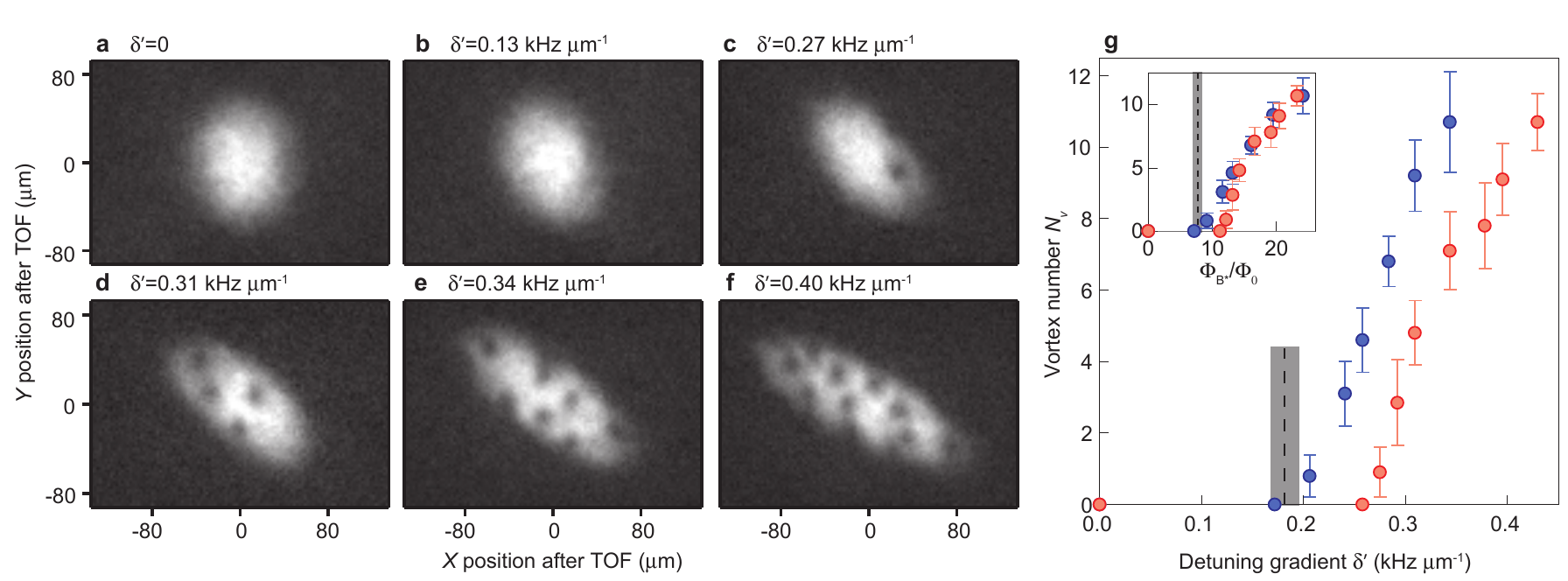}
\end{center}
\caption{{\bf Appearance of vortices at different detuning
gradients.} Data was taken for $N=1.4 \times 10^5$ atoms at hold
time $t_h=0.57$~s. {\bf a}-{\bf f}, Images of the $|m_F=0\rangle$
component of the dressed state after a 25.1~ms TOF with detuning
gradient $\delta^{'}$ from 0 to 0.43~kHz$\ \mu$m$^{-1}$ at Raman
coupling $\hbar \Omega_{R}=8.20$~$E_L$. {\bf g}, Vortex number $N_v$
versus $\delta^{'}$ at $\hbar \Omega_{R}=5.85$~$E_L$ (blue circles),
and $8.20$~$E_L$ (red circles). Each data point is averaged over
$\geq 20$ experimental realizations, and the uncertainties represent
standard deviations ($\sigma$). The inset displays $N_v$ versus the
synthetic magnetic flux $\Phi_{B^*}/\Phi_0={\cal A} q^*\langle
B^*\rangle /h$ in the BEC. The dashed lines indicate $\delta^{'}$
below which vortices become energetically unfavorable according to
our GPE computation, and the shaded regions show the 1-$\sigma$
uncertainty from experimental parameters.} \label{gradient}
\end{figure*}

We create a $^{87}$Rb BEC in a 1064~nm crossed dipole trap, loaded
into the lowest energy dressed state\cite{lin09prl} with atom number
$N$ up to $2.5\times 10^5$, and a Zeeman shift
$\omega_{Z}/2\pi=g\mu_{B}B/h \approx 2.71$~MHz, produced by a
\emph{real} magnetic bias field $B\hat{y}$. The $\lambda=801.7$~nm
Raman beams propagate along $\hat{y}\pm \hat{x}$ and differ in
frequency by a constant $\Delta \omega_L \simeq \omega_{Z}$, where a
small Raman detuning $\delta=\Delta \omega_{L}-\omega_{Z}$ largely
determines the vector potential $A^*_x$. The scalar light shift from
the Raman beams, combined with the dipole trap gives an
approximately symmetric three-dimensional potential, with
frequencies $f_x,f_y,f_z\approx 70$~Hz. Here, $\hbar k_L=h/(\sqrt
2\lambda)$ and $E_{L}=\hbar^2k_{L}^2/2m$ are the appropriate units
for the momentum and energy.

The spin and momentum states $|m_F,k_x\rangle$ coupled by the Raman
beams can be grouped into families of states labeled by the momentum
$\hbar k_x$. Each family
$\Psi(k_x)=\{|-1$,$k_x+2k_L\rangle$,$|0,k_x\rangle$,$|+1,k_x-2k_L\rangle
\}$ is composed of states that differ in linear momentum along
$\hat{x}$ by $\pm 2\hbar k_L$, and are Raman-coupled with strength
$\hbar \Omega_{R}$. For each $k_x$, the three dressed states are the
eigenstates in the presence of the Raman coupling, with energies
$E_j(k_x)$\cite{lin09prl}. The resulting vector potential is tunable
within the range $-2k_{L}<q^*A_{x}^*/\hbar<2k_{L}$. In addition,
$E_j(k_x)$ includes a scalar potential $V^{'}$\cite{Spielman09}.
$A_x^*$, $V^{'}$, and $m^*$ are functions of Raman coupling
$\Omega_{R}$ and detuning $\delta$, and for our typical parameters
$m^*\approx 2.5m$, reducing $f_x$ from $\approx 70$~Hz to $\approx
40$~Hz. The BEC's chemical potential $\mu/h \approx 1$~kHz is much
smaller than the $\sim h\times 10$~kHz energy separation between
dressed states, therefore the BEC only occupies the lowest energy
dressed state. Further, it justifies the harmonic expansion around
$q^*A_{x}^*/\hbar$, valid at low energy. Hence, the complete
single-atom Hamiltonian is ${\cal H}={\cal
H}^*_x+\hbar^2(k_y^2+k_z^2)/2m+V({\bf r})$, where $V({\bf r})$ is
the external potential including $V^{'}(\Omega_{R},\delta)$.

The dressed BEC starts in a uniform bias field ${\bf B}=B_0\hat{y}$,
at Raman resonance ($\delta=0$), corresponding to
$A^*_{x}=0$\cite{lin09prl}. To create a synthetic field $B^*$, we
apply a field gradient $b^{'}$ such that ${\bf
B}=(B_0-b^{'}y)\hat{y}$, ramping in 0.3 s from $b^{'}=0$ to a
variable value up to 0.055~Tm$^{-1}$, and then hold for $t_h$ to
allow the system to equilibrate. The detuning gradient $\delta^{'}=g
\mu_{B}b^{'}/\hbar$ generates a spatial gradient in $A_{x}^*$. For
the detuning range in our experiment, $\partial A_{x}^*/\partial
\delta$ is approximately constant, leading to an approximately
uniform synthetic field $B^{*}$ given by $B^*=\partial
A_{x}^*/\partial y=\delta^{'}\partial A_{x}^*/\partial \delta$
(Fig.~1d). To probe the dressed state, we switch off the dipole trap
and the Raman beams in less than $1\ \mu$s, projecting each atom
into spin and momentum components. We absorption-image the atoms
after a time-of-flight (TOF) ranging from 10.1~ms to 30.1~ms
(Fig.~1ef).

\begin{figure*}
\begin{center}
\includegraphics[width=160 mm]{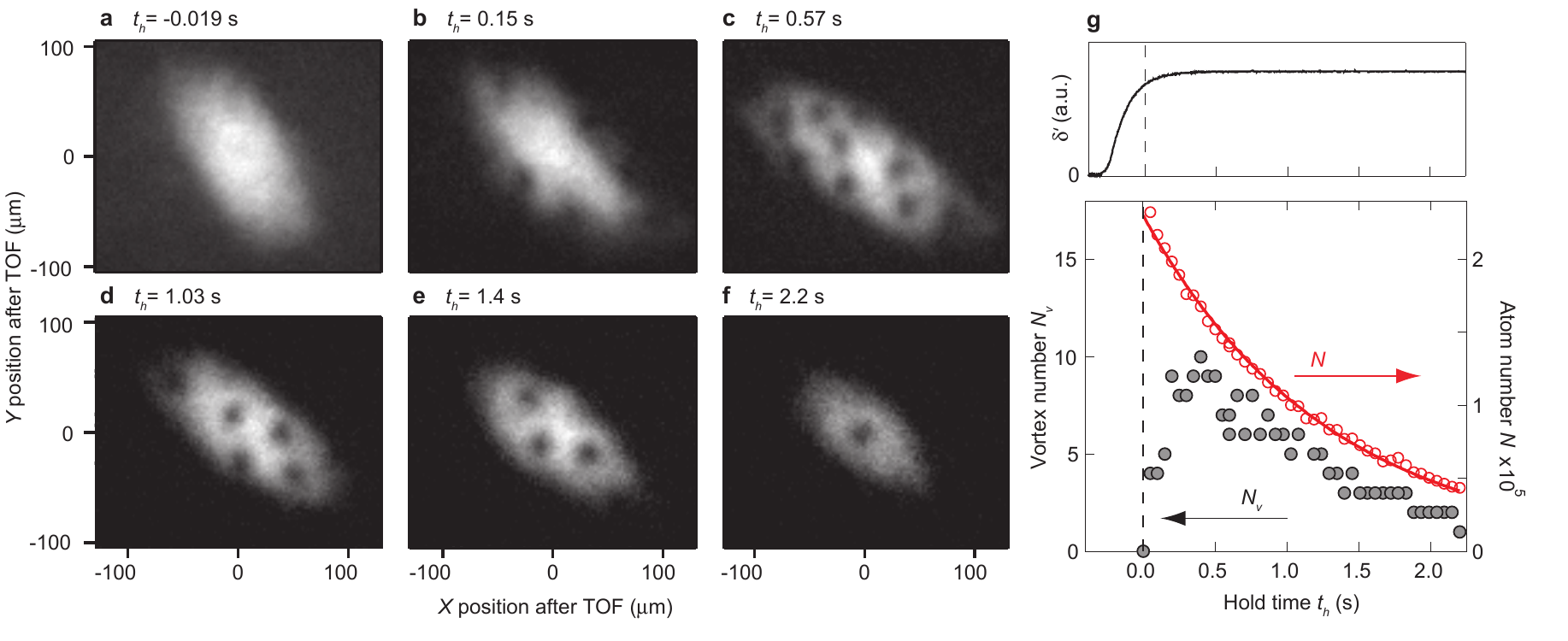}
\end{center}
\caption{{\bf Vortex formation.} {\bf a}-{\bf f} images of the
$|m_F=0\rangle$ component of the dressed state after a 30.1~ms TOF
for hold times $t_h$ between -0.019~s and 2.2~s. The detuning
gradient $\delta^{'}/2\pi$ is ramped to $0.31$~kHz$\ \mu$m$^{-1}$ at
the coupling $\hbar \Omega_{R}=5.85$~$E_L$. {\bf g}, top: time
sequence of $\delta^{'}$. bottom: vortex number $N_v$ (solid
symbols) and atom number $N$ (open symbols) versus $t_h$ with a
population lifetime of 1.4(2)~s. The number in parenthesis is the
uncorrelated combination of statistical and systematic 1-$\sigma$
uncertainties.} \label{holdtime}
\end{figure*}

\begin{figure}
\begin{center}
\includegraphics[width=80 mm]{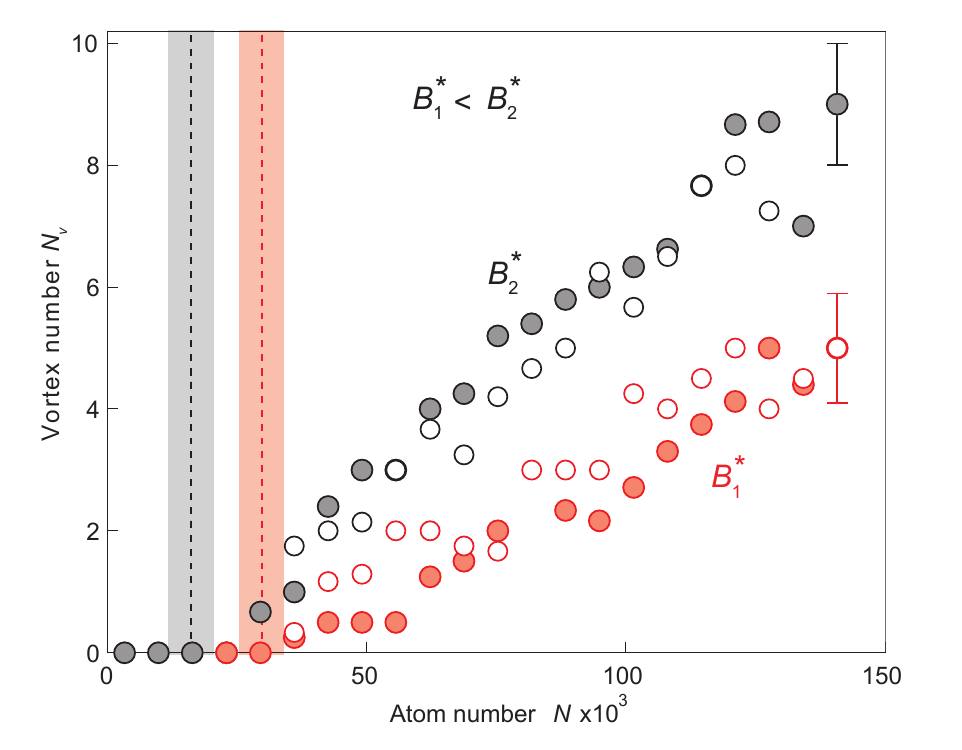}
\end{center}
\caption{{\bf Equilibrium vortex number.} Vortex number $N_v$ versus
atom number $N$ at detuning gradient $\delta^{'}_1/2\pi=0.26$~kHz$\
\mu$m$^{-1}$ (red circles) and $\delta^{'}_2/2\pi=0.31$~kHz$\
\mu$m$^{-1}$ (black circles), corresponding to synthetic fields
$B^*_1<B^*_2$, at Raman coupling $\hbar \Omega_{R}=5.85$~$E_L$. The
two data points with the largest $N$ show representative 1-$\sigma$
uncertainties, estimated from data in Fig.~2{\bf g}. We vary $N$ by
its initial value with a fixed hold time $t_h=0.57$~s (solid
symbols), and by $t_h$ with a fixed initial $N$ (open symbols). The
vertical dashed lines indicate $N$ below which vortices become
energetically unfavorable computed using our GPE simulation. The
shaded regions reflect the 1-$\sigma$ uncertainties from the
experimental parameters.} \label{atomnum}
\end{figure}

For a dilute BEC in low synthetic fields, we expect to observe
vortices. In this regime, the BEC is described by a macroscopic wave
function $\psi({\bf r})=|\psi({\bf r})|e^{i\phi({\bf r})}$, which
obeys the Gross-Pitaevskii equation (GPE). The phase $\phi$ winds by
$2\pi$ around each vortex, with amplitude $|\psi|=0$ at the vortex
center. The magnetic flux $\Phi_{B^*}$ results in $N_v$ vortices and
for an infinite, zero temperature system, the vortices are arrayed
in a lattice\cite{Yarmchuk79} with density $q^*B^*/h$. For finite
systems vortices are energetically less favorable, and their areal
density is below this asymptotic value, decreasing to zero at a
critical field $B^*_c$. For a cylindrically symmetric BEC, $B^*_c$
is given by $q^*B^*_{c}/h=5/(2\pi R^2)$ln$(0.67R/\xi)$ where $R$ is
the Thomas-Fermi radius and $\xi$ is the healing
length\cite{Lundh97}. $B^*_c$ is larger for smaller systems. For our
non-cylindrically symmetric system, we numerically solve the GPE to
determine $B^*_c$ for our experimental parameters (see Methods).

For synthetic fields greater than the critical value, we observe
vortices that enter the condensate and reach an equilibrium vortex
number $N_v$ after $\approx0.5$~s. Due to a shear force along
$\hat{x}$ when the Raman beams are turned off, the nearly-symmetric
insitu atom cloud tilts during TOF. While the vortices' positions
may rearrange, any initial order is not lost. During the time of our
experiment, the vortices do not form a lattice and the positions of
the vortices are irreproducible between different experimental
realizations, consistent with our GPE simulations. We measure
$N_{v}$ as a function of detuning gradient $\delta^{'}$ at two
couplings, $\hbar \Omega_{R}=5.85\ E_L$ and $8.20\ E_L$ (Fig.~2).
For each $\Omega_{R}$, vortices appear above a minimum gradient when
the corresponding field $\langle B^* \rangle=\delta^{'} \langle
\partial A_{x}^*/\partial \delta \rangle$ exceeds the critical field $B^*_c$.
(For our coupling $B^*$ is only approximately uniform over the
system and $\langle B^*\rangle$ is the field averaged over the area
of the BEC.) The inset shows $N_v$ for both values of $\Omega_R$
plotted versus $\Phi_{B^*}/\Phi_0={\cal A} q^*\langle B^*\rangle/h$,
the vortex number for a system of area ${\cal A}=\pi R_x R_y$ with
the asymptotic vortex density, where $R_x$ ($R_y$) is the
Thomas-Fermi radius along $\hat{x}\ $(or $\hat{y})$. Since the
system size, and thus $B^*_c$, are approximately independent of
$\Omega_{R}$, we expect this plot to be nearly independent of Raman
coupling. Indeed, the data for $\hbar \Omega_R=5.85 E_L$ and $8.20
E_L$ only deviate for $N_v < 5$, likely due to the intricate
dynamics of vortex nucleation\cite{Murray2009}.

Figure~3 illustrates a progression of images showing vortices
nucleate at the system's edge, fully enter to an equilibrium density
and then decay along with the atom number. The time scale for vortex
nucleation depends weakly on $B^{*}$, and is more rapid for larger
$B^*$ with more vortices: It is $\approx 0.3$~s for vortex number
$N_v \geq 8$, and increases to $\approx 0.5$~s for $N_v=3$. For
$N_v=1$ ($B^*$ near $B^*_c$), the single vortex always remains near
the edge of the BEC. In the dressed state, spontaneous emission from
the Raman beams removes atoms from the trap, causing the population
to decay with a 1.4(2)~s lifetime, and the equilibrium vortex number
decreases along with the BEC's area.

To verify the dressed state has reached equilibrium, we prepare
nominally identical systems in two different ways. First, we vary
the initial atom number and measure $N_v$ as a function of atom
number $N$ at a fixed hold time $t_{h}=0.57$~s. Second, starting
with a large atom number, we measure both $N_v$ and $N$, as they
decrease with $t_h$ (Fig.~3). Figure~4 compares $N_v$ versus $N$
measured with both methods, each at two detuning gradients
corresponding to fields $B^*_1<B^*_2$. The data show $N_v$ as a
function of $N$ is the same for these preparation methods, providing
evidence that for $t_h \geq 0.57$~s, $N_v$ has reached equilibrium.
As the atom number $N$ falls, the last vortex departs the system
when the critical field -- increasing with decreasing $N$ --
surpasses the actual field.

In conclusion, we have demonstrated optically synthesized magnetic
fields for neutral atoms resulting from the Berry's phase, a
fundamental concept in physics. This novel approach differs from
experiments with rotating gases, where it is difficult to add
optical lattices and rotation is limited by heating, metastability,
and the difficulties to add large angular momentum, preventing
access to the quantum-Hall regime. A standout feature in our
approach is the ease to add optical lattices. For example, the
addition of a 2D lattice makes it immediately feasible to study the
fractal energy levels of the Hofstadter
butterfly\cite{Hofstadter76}. Further, a 1D lattice can divide the
BEC into an array of 2D systems normal to the field. A suitable
lattice configuration allows access to the $\nu\sim 1$ quantum-Hall
regime, with an ensemble of 2D systems each with $\approx 200$
atoms, and with a realistic $\approx k_{B}\times 20$~nK interaction
energy.

We thank W.D. Phillips for discussions.  This work was partially
supported by ONR, ARO with funds from the DARPA OLE program, and the
NSF through the JQI Physics Frontier Center. R.L.C. acknowledges the
NIST/NRC postdoctoral program and K.J.G. thanks CONACYT.

\begin{center}{\bf Methods Summary}
\end{center}
\subsection{Dressed state preparation}
We create a $^{87}$Rb BEC in a crossed dipole trap\cite{Lin09}, with
$N\approx 4.7 \times 10^5$ atoms in $|F=1,m_{F}=-1\rangle$. The
quadratic Zeeman shift is $\hbar \epsilon=0.61 E_L$ for
$\omega_Z/2\pi=g\mu_{B}B/h \approx 2.71$~MHz, where $g$ is the
Land$\acute{\rm e}$ $g$-factor. To maintain $\delta=0$ at the BEC's
center as we ramp the field gradient $b^{'}$, we change
$g\mu_{B}B_0$ by as much as $7 E_L$. Simultaneously, we decrease the
dipole beam power by $20\%$, producing our $\approx 40$~Hz trap
frequency along $\hat{x}$. Additionally, the detuning gradient
$\delta^{'}\hat{y}$ makes the scalar potential $V^{'}$ anti-trapping
along $\hat{y}$, reducing $f_y$ from 70~Hz to 50~Hz for our largest
$\delta^{'}$.  Spontaneous emission from the Raman beams decreases
the atom number to $N\approx 2.5\times 10^5$ for $t_h=0$, with a
condensate fraction of $0.85$.

\subsection{Numerical method} We compare our data to a
finite temperature 2D stochastic GPE (SGPE\cite{blakie08})
simulation including the dressed state dispersion $E(k_x,y)$ that
depends on $y$ through the detuning gradient $\delta^\prime$. We
evolve the time-dependent projected GPE
\begin{widetext}
$$
i\hbar\frac{\partial\psi({\bf x},t)}{\partial t}={\mathcal
P}\left\{\left[E\left(-i\hbar\frac{\partial}{\partial
x},y\right)-\frac{\hbar^2}{2 m}\frac{\partial^2}{\partial
y^2}+g_{2D}\left|\psi({\bf x},t)\right|^2\right]\psi({\bf
x},t)\right\}.
$$
\end{widetext} $\mathcal P$ projects onto a set of significantly
occupied modes, and $g_{2D}$ parameterizes the 2D interaction
strength.  The SGPE models interactions between the highly-occupied
modes described by $\psi$ and sparsely occupied thermal modes with
dissipation and an associated noise term. We approximately account
for the finite extent along $\hat z$ by making $g_{2D}$ depend on
the local 2D density . For low temperatures this 2D model correctly
recovers the 3D Thomas-Fermi radii, and gives the expected 2D
density profile. These quantitative details are required to
correctly compute the critical field or number for the first vortex
to enter the system, which are directly tied to the 2D condensate
area.


\end{document}